%% file: cascella.tex
\documentclass[11pt]{article}
\usepackage[latin9]{inputenc}
\usepackage{amsmath}
\usepackage{amssymb}
\usepackage{graphicx}

\makeatletter
\usepackage{xspace}
\usepackage{color}

\input econfmacros.tex
\def\Title#1{\begin{center} {\Large {\bf #1} } \end{center}}

\makeatother

\begin{document}
\Title{The SuperNEMO tracking detector}

\bigskip{}
\bigskip{}


\begin{raggedright}
\textit{Michele Cascella\index{Cascella, M.},}\\
\textit{Department of Physics and Astronomy}\\
\textit{University College London}\\
\textit{Gower Street, WC1E 6BT, London, UK }\\

\end{raggedright} \vspace{1cm}

\begin{flushleft}
\emph{\small{}To appear in the proceedings of the Prospects in Neutrino
Physics Conference, 15 -- 17 December, 2014, held at Queen Mary University
of London, UK.}{\small{} }
\par\end{flushleft}{\small \par}

\section{Introduction}

The SuperNEMO detector \cite{SuperNEMO} will search for neutrinoless
double beta decay at the Modane Underground Laboratory on the French-Italian
border. This decay mode, if observed, would be proof that the neutrino
is its own antiparticle, would constitute evidence for total lepton
number violation, and could allow a measurement of the absolute neutrino
mass. 

The SuperNEMO experiment is conceived as 20 identical planar modules,
each contains 5-7 kg of $\beta\beta$ isotope (the baseline isotope
is $^{82}\mbox{Se}$ but $^{150}\mbox{Nd}$ and $^{48}\mbox{Ca}$
are also candidates); it is designed to reach a half-life sensitivity
of $10^{26}$  years corresponding to an effective Majorana neutrino
mass of $50-100$ meV. 

The fist module of the SuperNEMO detector is currently under construction
and the first quarter of the tracking detector has recently been completed
and is being commissioned.

\section{Neutrinoless Double Beta Decay}

Double beta decay is a rare but well-understood process. It has been
observed in 11 nuclei for which the simultaneous $\beta$ decay of
two nucleons is an energetically advantageous process. The neutrino-less
($0\nu$) version of this process is potentially a window to observe
two Majorana neutrinos annihilating with each other. 

Neutrinos are the only neutral fermions, their masses are much lighter
than any other massive particles. Ettore Majorana observed that if
the neutrino is truly neutral, it is possible to write a Lagrangian
in which the neutrino is its own antiparticle. The Majorana mass term
is very different from the Dirac term that describes all other fermions,
allowing for the neutrino mass to be generated with only the left
handed antineutrinos and right handed antineutrinos, and violating
absolute lepton number conservation. 

The Majorana mass generation mechanism enables the creation of see-saw
models. These models explain the unnatural lightness of neutrinos
through the introduction of additional heavy neutrinos at the GUT
scale. If proven, these models would offer a way of probing GUT-scale
physics at accessible energies.

\section{The SuperNEMO detector and the Demonstrator Module}

\begin{figure}
\begin{centering}
\includegraphics[height=0.15\textheight]{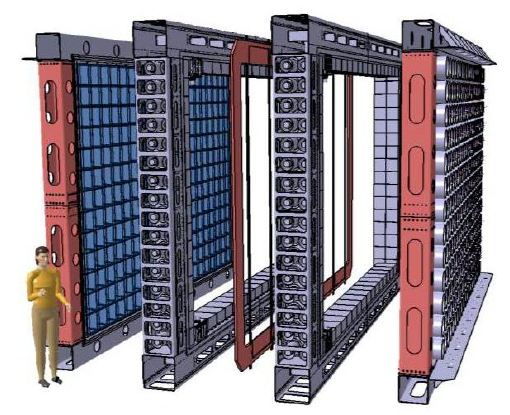}\includegraphics[height=0.15\textheight]{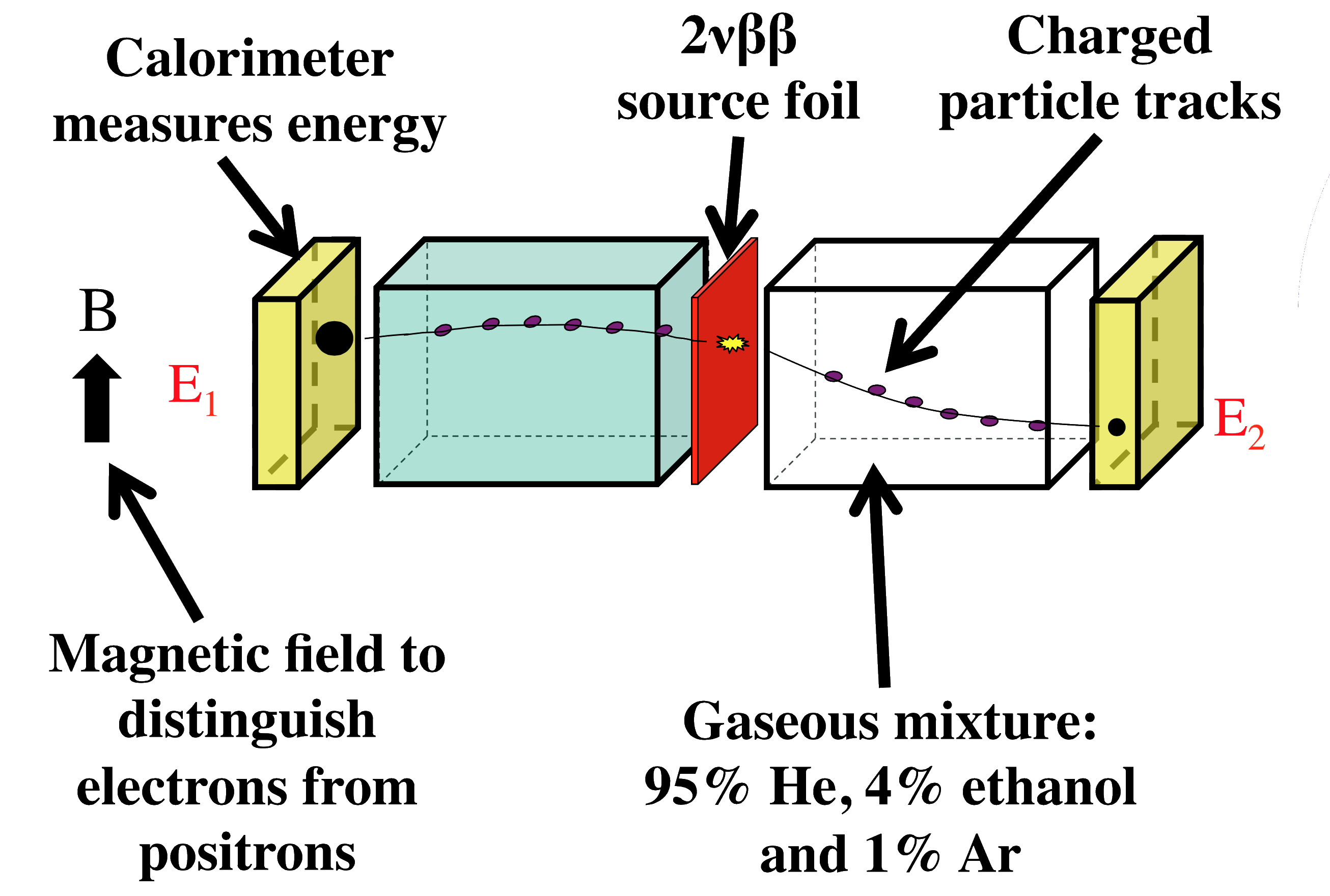}
\par\end{centering}

\protect\caption{\label{fig:A-SuperNEMO-module}A SuperNEMO module (left), and a scheme
of the detector geometry (right). This geometry allows full event
topology reconstruction ($e^{\pm}/\alpha/\gamma$ separation) and
the freedom to choose the source element. }

\end{figure}

The main feature that separates the SuperNEMO experiment from the
other $0\nu2\beta$ experiments is the fact that the source element
is completely decoupled from the detector. Building on the NEMO3 experience
\cite{NEMO3} a SuperNEMO module (shown in figure \ref{fig:A-SuperNEMO-module})
has an electron tracker and a calorimeter, allowing for the complete
reconstruction of the decay event topology. This allows unprecedented
levels of background rejection: electrons can be effectively separated
from positrons, $\alpha$ and $\gamma$ particles. 

Great attention has also been given to material selection and screening
in order to achieve unprecedented radiopurity. For the same reason
every step of the construction and assembly of the detector happens
in a cleanroom environment. Thanks to these efforts the expected background
in the region of interest is $10^{-4}$ events/keV/kg/yr. 

This detector design also allows complete topological reconstruction
of the double beta decay event; in the event of a discovery, such
topological measurements will be essential in determining the nature
of the lepton number violating process. 

The first module, called the Demonstrator Module, is currently under
construction whit the aim of proving the feasibility of the full experiment.
This module will host 7~kg of $^{82}$Se, and has an expected sensitivity
of $T_{1/2}^{0\nu}>6.6\times10^{24}$~y (corresponding to $|m_{\beta\beta}|<0.2-0.4$~eV)
 after 2.5~y.

\section{The Demonstrator tracker}

The Demonstrator Module tracker has 2034 2~m long octagonal drift
cells operating in Geiger mode in a He/ethanol/Ar (95:4:1) mixture. 

In NEMO3 the main background for the $0\nu2\beta$ search is the radon
inside the tracker volume whose electrons have enough energy to mimic
a double-$\beta$ event. For this reason only selected materials (copper,
steel, Duracon and PTFE) are allowed, the construction process and
most components of the tracker have been screened with ultra sensitive
Rn detectors \cite{Xin}. For the same reason the tracker gas is also
purified with a Rn trap using cold carbon filters. 

\begin{figure}
\begin{centering}
\includegraphics[width=0.6\textwidth]{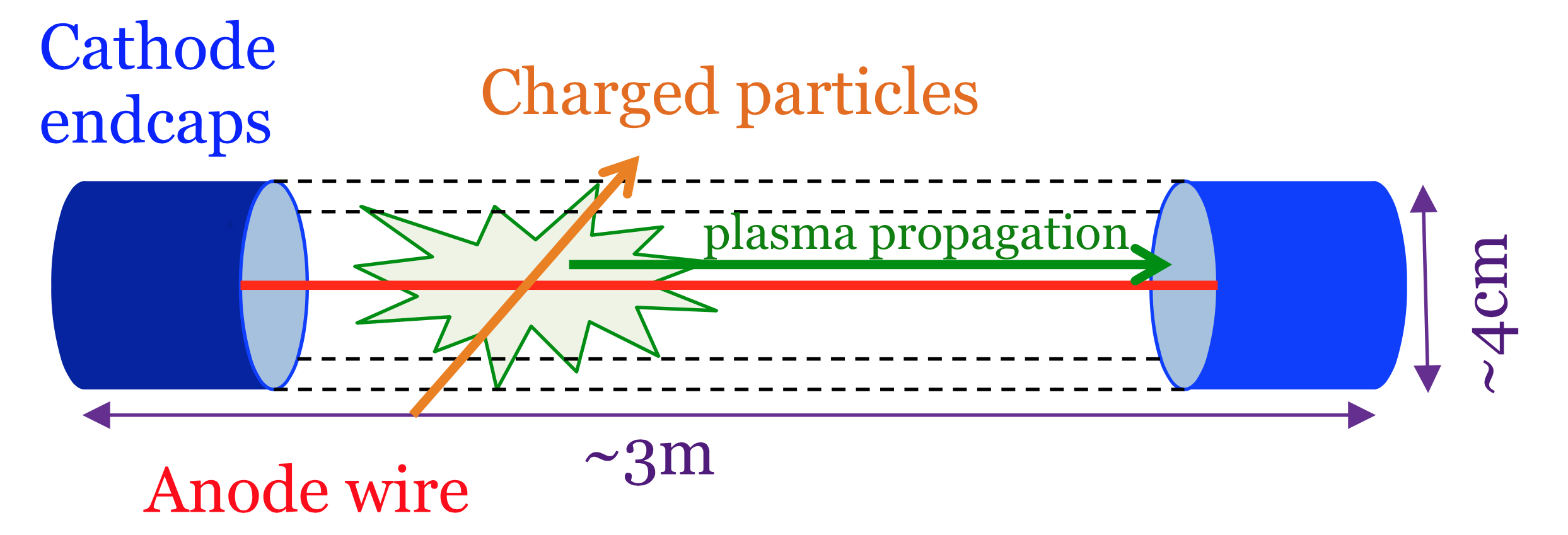}\includegraphics[width=0.33\textwidth]{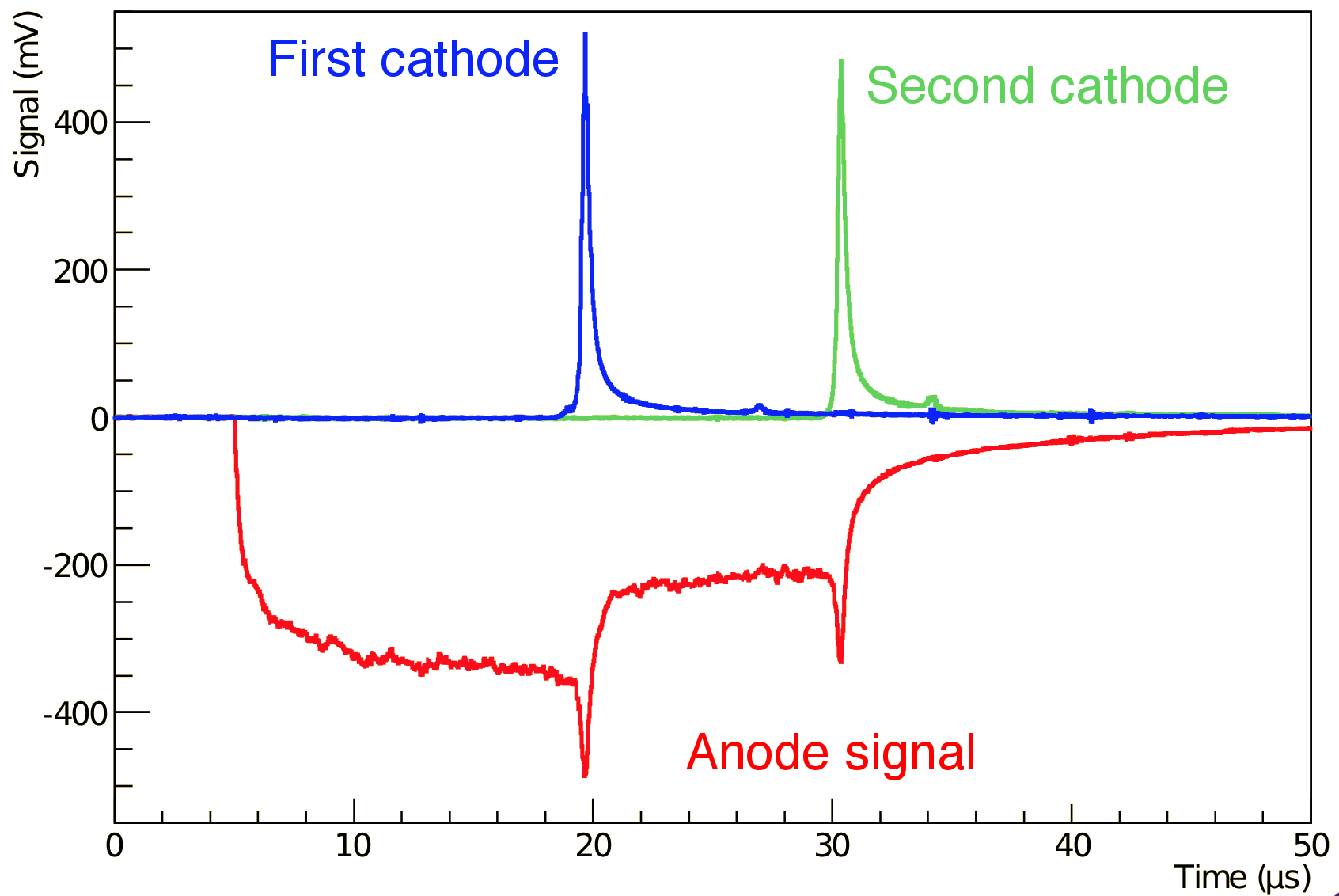}
\par\end{centering}

\protect\caption{\label{fig:A-SuperNEMO-tracker-cell}A SuperNEMO tracker cell (left)
and a depiction of the electrical signal (right) measured for the
anode (negative) and cathodes (positive).}
\end{figure}

The distance of closest approach of the track with respect to the
anode wire is measured from the time of the anode signal; the time
difference between the Geiger discharge arriving at each end of the
cell provides the longitudinal location of the track (see figure \ref{fig:A-SuperNEMO-tracker-cell}).

\paragraph{Cell production}

\begin{figure}
\begin{centering}
\includegraphics[height=0.15\textheight]{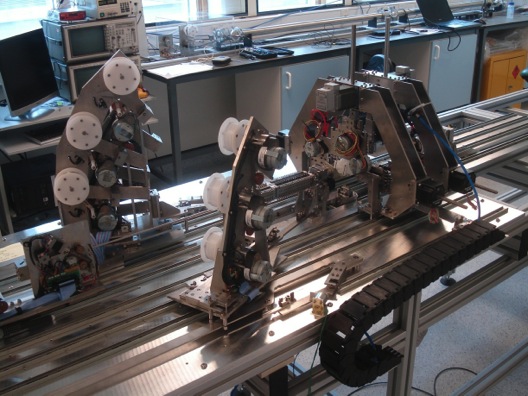} \includegraphics[height=0.15\textheight]{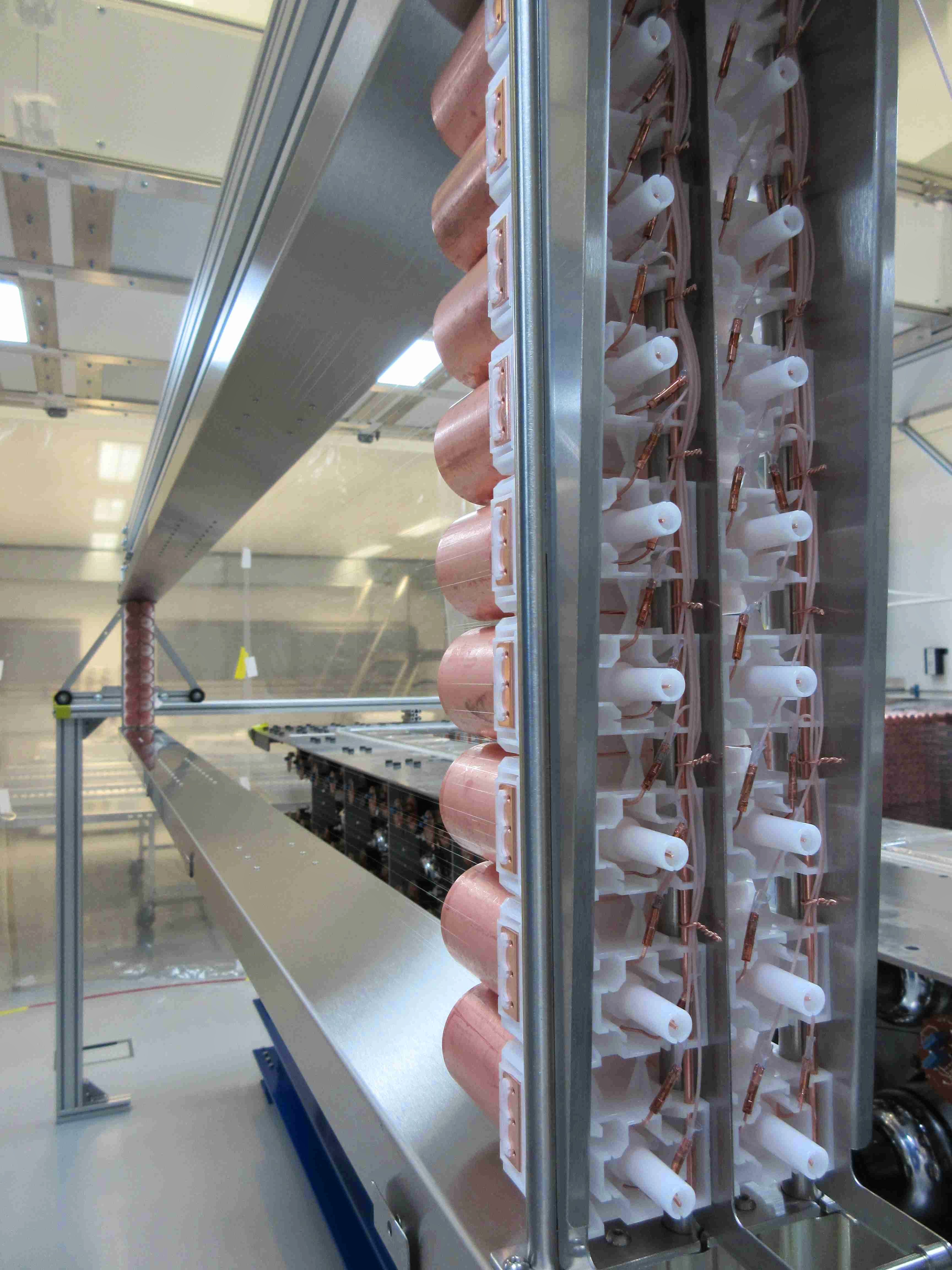}
\includegraphics[height=0.15\textheight]{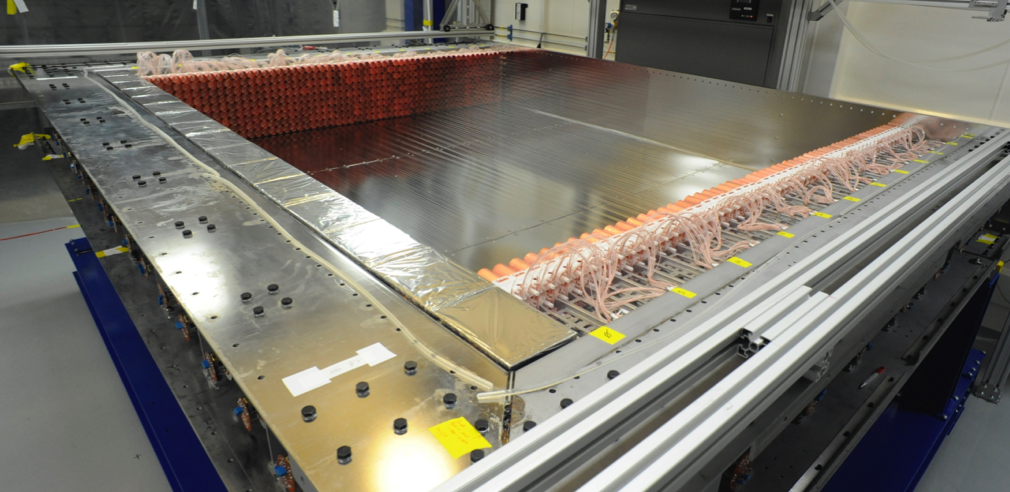}
\par\end{centering}

\protect\caption{\label{fig:The-wiring-robot}The wiring robot (left), a cassette being
inspected before installation (center), and the \label{fig:first-section}first
of the four sections that make up the Demonstrator Module (right). }
\end{figure}

To ensure the radiopurity goal is not compromised all tracker components
are cleaned, and, where needed, passivated to prevent corrosion. To
minimize the contamination, the drift cells are assembled by a wiring
robot (pictured in figure \ref{fig:The-wiring-robot}), in a cleanroom
environment at the University of Manchester. 

The fundamental production unit is a cassette made of 2 columns of
9 drift cells. Each cassette is conditioned immediately after production
to eliminate self-triggering or plasma-blocking points that can be
caused by the presence of small impurities on the wires. The conditioning
is prolonged until a wide enough Geiger plateau is observed on all
cells.

\paragraph{Tracker assembly and commissioning}

The tracking detector is divided in 4 sections each containing 56
rows of drift cells (28 cassettes). After conditioning the cassettes
are transferred to the Mullard Space Science Laboratory where a clean
tent large enough to host the a whole section of the tracker has been
erected. There each cassette is tested, inspected, installed on the
tracker frame and connected to the feedthroughs that transport the
electrical signals outside the gas volume. 

Each section has been designed so that it can be sealed, tested and
shipped to the LSM independently. The first SuperNEMO section has
already been assembled, sealed and is now ready to for surface commissioning;
a picture of the first tracker section just before sealing is shown
in figure \ref{fig:first-section}. 

\begin{figure}
\begin{centering}
\includegraphics[height=0.2\textheight]{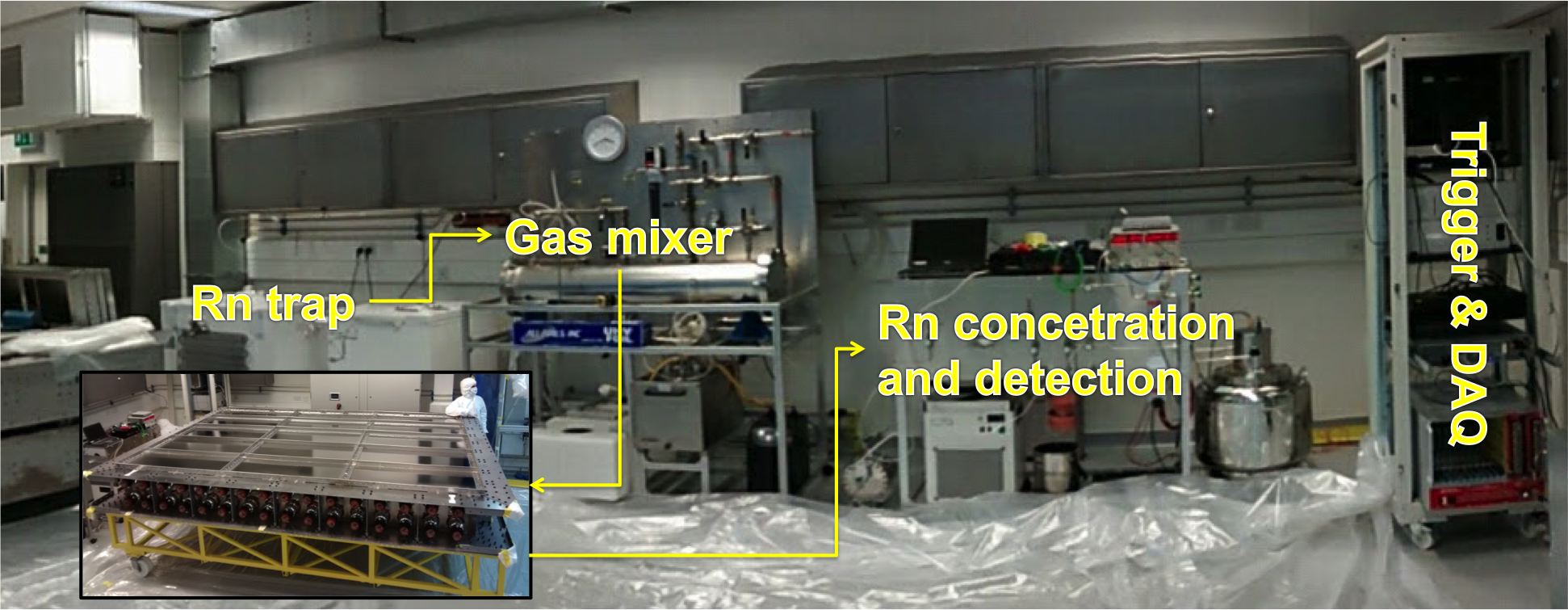} 
\par\end{centering}

\protect\caption{\label{fig:The-commissioning-system;}The commissioning system; the
arrows show the path of the gas that is filtered by the Rn trap, flown
trough the tracker section and then passed to the Rn concentration
line to measure the tracker intrinsic emanation.}
\end{figure}

The Rn emanated by the tracker will be monitored for several weeks
to make sure that the background is within the requirements (see figure
\ref{fig:The-commissioning-system;} as well as \cite{Xin}). After
Rn testing the tracker will be powered in stages using refurbished
NEMO3 electronics and commissioned using cosmic rays. While the the
first section is tested for radio-purity and commissioned the frame
for the second section to be built, is being assembled.

\section{Summary}

SuperNEMO is a detector to search for neutrinoless double beta decay
mode in one or more isotopes. To reach an extremely low background
rate in the signal region the experiment has a separate tracking and
calorimetric section; great care is also devoted to the material selection,
the cleanliness of the construction and assembly chain, and the testing
of each component.

The Demonstrator Module is currently under construction and is scheduled
to be completed by the end of 2015; the commissioning will begin in
2016 in Modane.

\end{document}

%% file: econfmacros.tex
\definecolor{Red}{rgb}{1,0,0}
\definecolor{Green}{rgb}{0,1,0}
\definecolor{Blue}{rgb}{0,0,1}
\definecolor{Black}{rgb}{0,0,0}

\def\beq{\begin{equation}}
\def\eeq#1{\label{#1}\end{equation}}
\def\eeqn{\end{equation}}

\def\beqa{\begin{eqnarray}}
\def\eeqa#1{\label{#1}\end{eqnarray}}
\def\eeqan{\end{eqnarray}}

\let\bar=\overbar

\def\Dslash{\not{\hbox{\kern-4pt $D$}}}
\def\dslash{\not{\hbox{\kern-2pt $\del$}}}

\def\msb{{\bar{\ssstyle M \kern -1pt S}}}